\documentclass[12pt,preprint]{aastex}

\shorttitle{Modeling Coronal Loops}
\shortauthors{Warren et al.}

\begin{document}


\title{Modeling Evolving Coronal Loops with Observations from \textit{STEREO},
  \textit{Hinode}, and \textit{TRACE}}

\author{%
  Harry P. Warren,
  David M. Kim\altaffilmark{1}, 
  Amanda M. DeGiorgi\altaffilmark{2}, 
  Ignacio Ugarte-Urra\altaffilmark{3}
}
\affil{ Space Science Division, Naval Research Laboratory, Washington, DC
  20375} 
\altaffiltext{1}{Thomas Jefferson High School for Science and Technology,
  Alexandria, VA} 
\altaffiltext{2}{University of Chicago, Chicago, IL}
\altaffiltext{3}{College of Science, George Mason University, 4400 University Drive,
  Fairfax, VA 22030}


\begin{abstract}
  The high densities, long lifetimes, and narrow emission measure distributions observed
  in coronal loops with apex temperatures near 1\,MK are difficult to reconcile with
  physical models of the solar atmosphere. It has been proposed that the observed loops
  are actually composed of sub-resolution ``threads'' that have been heated impulsively
  and are cooling. We apply this heating scenario to nearly simultaneous observations of
  an evolving post-flare loop arcade observed with the EUVI/\textit{STEREO},
  XRT/\textit{Hinode}, and \textit{TRACE} imagers and the EIS spectrometer on
  \textit{HINODE}. We find that it is possible to reproduce the extended loop lifetime,
  high electron density, and the narrow differential emission measure with a multi-thread
  hydrodynamic model provided that the time scale for the energy release is sufficiently
  short. The model, however, does not reproduce the evolution of the very high temperature
  emission observed with XRT. In XRT the emission appears diffuse and it may be that this
  discrepancy is simply due to the difficulty of isolating individual loops at these
  temperatures. This discrepancy may also reflect fundamental problems with our
  understanding of post-reconnection dynamics during the conductive cooling phase of loop
  evolution.
\end{abstract}



\section{Introduction}

One of the principal problems in solar physics is understanding how the Sun's corona is
heated to very high temperatures. Recent work on coronal loops indicates that they have
physical properties that are difficult to reconcile with theoretical models. Coronal loops
with temperatures near 1\,MK are observed to persist longer than a characteristic cooling
time, suggesting steady or quasi-steady heating
\cite[e.g.,][]{lenz1999,aschwanden2000b}. Steady heating models, however, cannot reproduce
the high electron densities observed in these loops \citep{winebarger2003}.  Multi-thread,
impulsive heating models have been proposed as a possible heating scenario
\cite[e.g.,][]{cargill1997,warren2003,patsourakos2008}.  Such models are motivated by our
understanding of the energy release during magnetic reconnection in flares
\citep[e.g.,][]{parker1983}. In these models impulsive heating leads to high densities and
multiple, sub-resolution ``threads'' lead to long lifetimes relative to the cooling time
for an individual loop. These models are severely constrained by the relatively narrow
distributions of temperatures that are often observed in loops with apex temperatures near
1\,MK \citep[e.g.,][]{delzanna2003,aschwanden2005b,cirtain2007,warren2008b}. A narrow
distribution of temperatures suggests that the loop can contain only a few independent
threads.

One difficulty with fully testing coronal heating scenarios such as these with
hydrodynamic models has been the spareness of data. Previous work on loop evolution has
generally focused on measurements imaging instruments
\cite[e.g.,][]{warren2003,aschwanden2005b,urra2006}, which have limited diagnostic
capabilities. Current solar observatories, however, allow for coronal loops to be observed
in unprecedented detail. The EUV Imaging Spectrometer (EIS) on the \textit{Hinode} mission
provides high spatial and spectral resolution observations over a very wide range of
coronal temperatures. EIS plasma diagnostics yield important constraints on the physical
properties of coronal loops. The X-ray Telescope (XRT) on \textit{Hinode} complements
these observations with high spatial and temporal resolution observations of the high
temperature corona. The multiple viewpoints of the twin \textit{STEREO} spacecraft allow
for loop geometry, a critical parameter in the modeling, to be measured using the EUV
Imagers (EUVI). The \textit{Transition Region and Coronal Explorer} (\textit{TRACE})
currently provides the highest spatial resolution images of the solar corona.

In this paper we use \textit{STEREO}, \textit{Hinode}, and \textit{TRACE} observations of
an evolving loop in a post-flare loop arcade to make quantitative comparisons between a
multi-thread, impulsive heating model and measured densities, temperatures, intensities
and loop lifetimes. An important component of this work is the development of methods for
integrating the different observations into hydrodynamic simulations of the loop. We find
that it is possible to reproduce the extended loop lifetime, the high electron density,
and the narrow differential emission measure (DEM) with a multi-thread model provided the
time scale for the energy release is sufficiently short. The model, however, does not
reproduce the evolution of the high temperature emission observed with XRT.

One goal of investigating the heating on individual loops is to motivate the modeling of
entire active regions or even the full Sun
\cite[e.g.,][]{schrijver2004,warren2006b,mok2005,lundquist2008}. It is possible, however,
that there is not a single coronal heating mechanism that can be applied to all coronal
loops. For example, it may be that steady heating is the dominant heating scenario on some
fraction of coronal loops \citep[e.g.,][]{martens2008,antiochos2003}. Even if impulsive
heating of the kind discussed here is only a minor contributor to the heating of the solar
corona, this study provides important insights into the energy release during magnetic
reconnection, a fundamental process in astrophysical and laboratory plasmas.

\section{Observations}

\begin{figure}
\centerline{%
  \includegraphics[clip,scale=0.56]{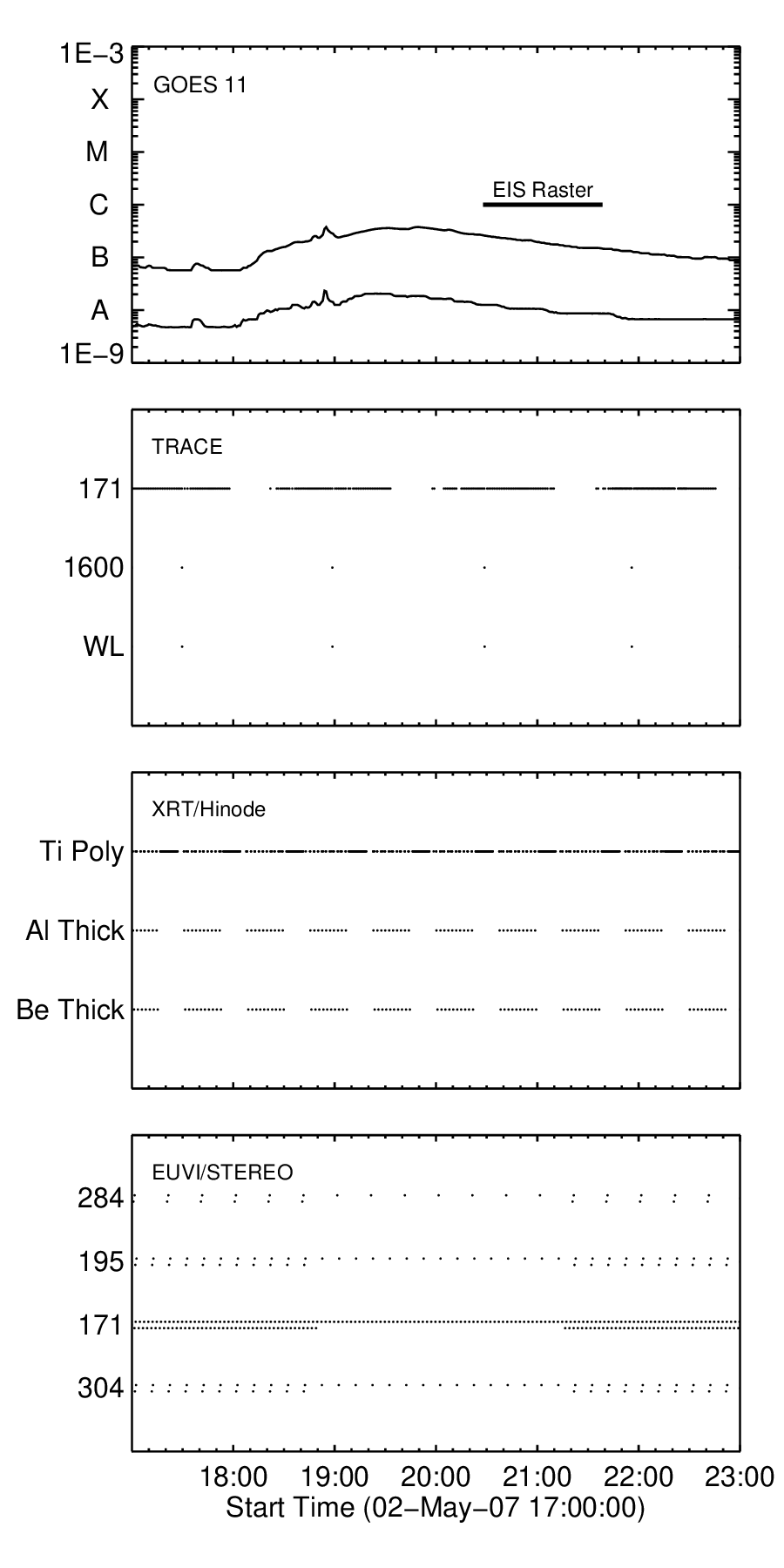}}
\caption{A summary of the GOES, EIS, TRACE, XRT, and EUVI data taken on May 2, 2007
  between 17:00 and 23:00 UT. Each image is indicated with a dot. For EUVI the times of
  the \textit{STEREO A} images are shown below the \textit{STEREO B} images. There are
  gaps in the TRACE data due to orbital eclipses. There is also a gap in the
  \textit{STEREO A} data.}
\label{fig:summary}
\end{figure}

In this section we provide an overview of the instruments and observations used in this
study. A summary of the observations is shown in Figure~\ref{fig:summary}. The loop
considered here is a post-flare loop from a very small event (GOES class B2.5) that peaked
around 19:00 UT on May 2, 2007.

The EIS instrument on \textit{Hinode} produces stigmatic spectra in two wavelength ranges
(171--212\,\AA\ and 245--291\,\AA) with a spectral resolution of 0.0223\,\AA. There are
1\arcsec\ and 2\arcsec\ slits as well as 40\arcsec\ and 266\arcsec\ slots available. The
slit-slot mechanism is 1024\arcsec\ long but a maximum of 512 pixels on the CCD can be
read out at one time. Solar images can be made using one of the slots or by stepping one
of the slits over a region of the Sun. Telemetry constraints generally limit the spatial
and spectral coverage of an observation. See \cite{culhane2007} and \cite{korendyke2006}
for more details on the EIS instrument.

\begin{figure*}
\centerline{%
  \includegraphics[clip,scale=1.00]{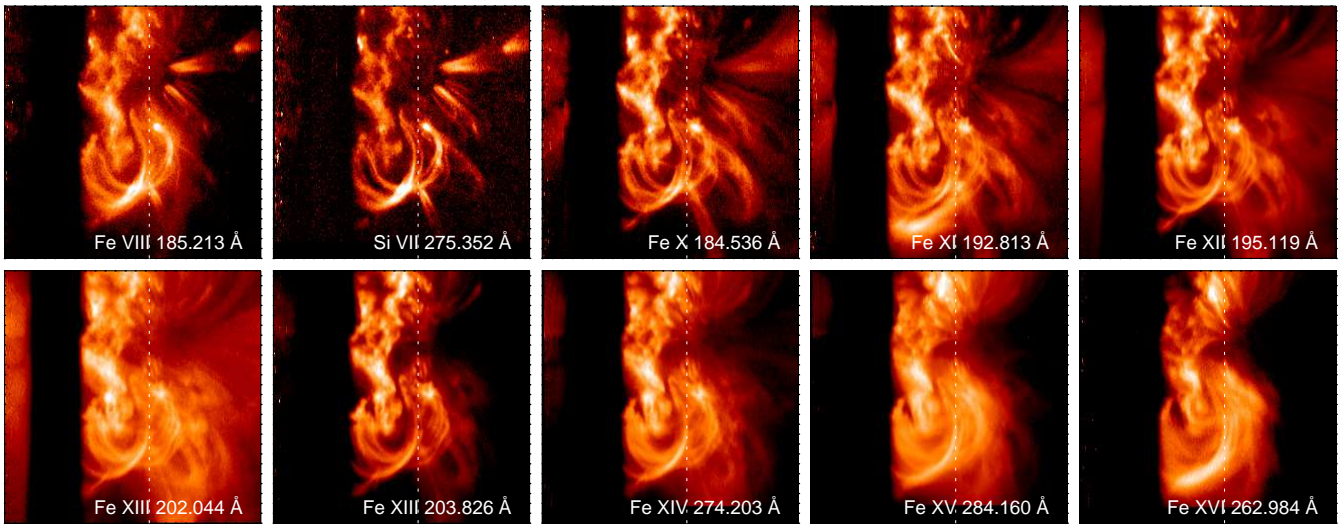}}
\caption{EIS rasters of active region 10953 in 10 different emission lines. These images
  were constructed by stepping the 1\arcsec\ slit from west (right) to east (left) over
  the region between 20:27 and 21:38 on May 2, 2007. The field of view is
  $256\arcsec\times256\arcsec$. Each exposure is 15\,s in duration.  Note that the
  \ion{Fe}{11} 192.813\,\AA\ raster shows both \ion{Fe}{11} and \ion{Ca}{17} 192.858\,\AA\
  emission. The dark vertical band is from a \textit{Hinode} orbital eclipse. For
  comparison with the imaging data a vertical line has been drawn on the rasters
  corresponding to data taken 20:58:34 UT.}
\label{fig:eis}
\end{figure*}

For these observations the 1\arcsec\ slit was stepped over the active region and 15\,s
exposures were taken at each position. An area of $256\arcsec\times256\arcsec$ was imaged
in about 71 minutes.  A total of 20 spectral windows were read out of the CCD and included
in the telemetry stream. The raw data were processed using \verb+eis_prep+ to remove dark
current, warm pixels, and other instrumental effects using standard software. During the
processing the observed count rates are converted to physical units. Intensities from the
processed data are computed by fitting the observed line profiles with Gaussians. The EIS
rasters are co-aligned to account for any spatial offsets (see \citealt{young2009b} for a
discussion). Spacecraft jitter during the raster has not been accounted for. The
\textit{Hinode} housekeeping logs suggest relatively small displacements (less than one
pixel) for the narrow field of view of interest here. For larger structures spacecraft
jitter can be important. EIS rasters in a number of different emission lines are shown in
Figure~\ref{fig:eis}, and show post-flare loops at various temperatures in the lower part
of the active region. These rasters also indicate a brief data gap due to orbital eclipse.

One limitation of these EIS data is the lack of temporal information. Better information
on the temporal evolution of these loops is provided by the imaging instruments, such as
the EUVI \citep{howard2008} on the \textit{Solar Terrestrial Relations Observatory}
(\textit{STEREO}) mission. The EUVI is a normal incidence, multilayer telescope which can
observe the Sun in 4 wavelength bands centered at 284, 195, 171, and 304\,\AA. EUVI
observes the full Sun and therefore has reduced spatial resolution (1.6\arcsec\ pixels)
relative to the other observations that we consider here. There are two \textit{STEREO}
spacecraft with identical instrument packages. The twin \textit{STEREO} spacecraft drift
away from the Earth at about 23$^\circ$ per year. On May 2, 2007 the separation between
the spacecraft was small, about 6$^\circ$.

\begin{figure*}
\centerline{%
  \includegraphics[clip,scale=1.00]{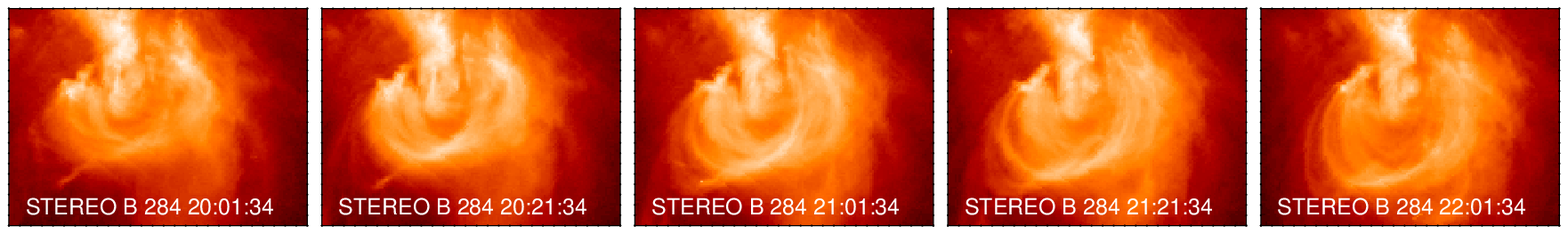}}
\centerline{%
  \includegraphics[clip,scale=1.00]{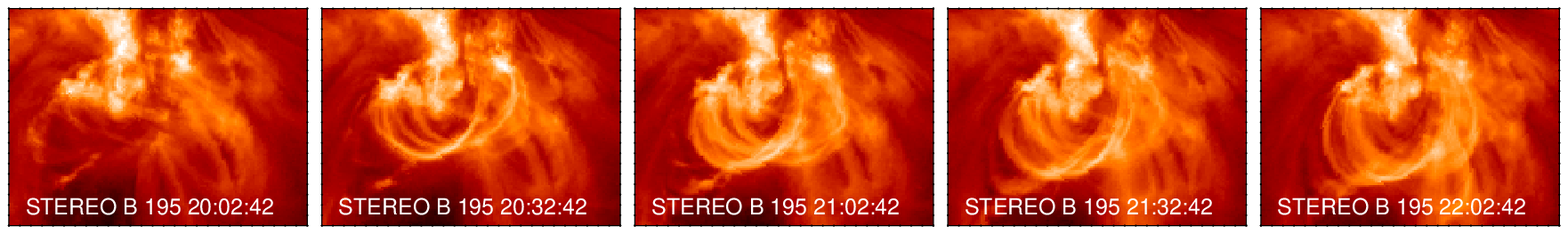}}
\centerline{%
  \includegraphics[clip,scale=1.00]{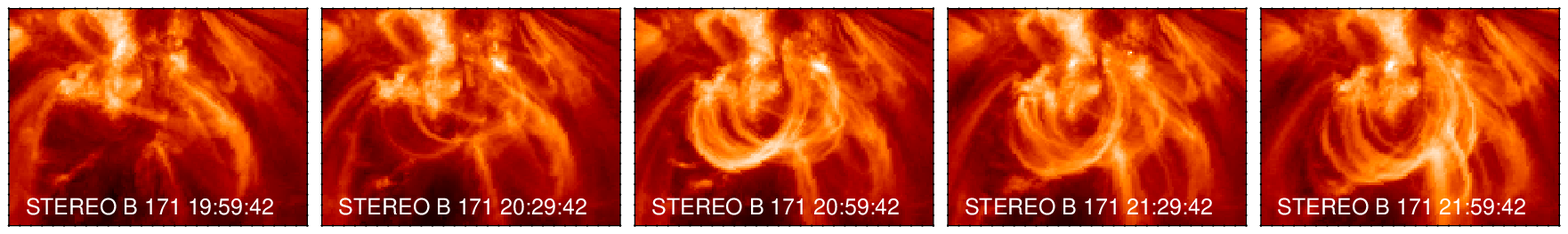}}
\caption{EUVI/\textit{STEREO B} 284, 195, and 171\,\AA\ images of active region 10953 from
  approximately 20 to 22 UT on May 2, 2007. The field of view is approximately
  $216\arcsec\times168\arcsec$.}
\label{fig:euvi}
\end{figure*}

The EUVI images taken around the time of the EIS raster are indicated in
Figure~\ref{fig:summary}. Because of telemetry constraints the image cadence is
limited. For these observations 171\,\AA\ images were taken at a relatively high cadence
($\sim 150$\,s) while the images at the other wavelengths were taken at lower cadences
($\sim 600$--1200\,s). The raw data are processed using \verb+euvi_prep+ to produce
calibrated, co-aligned images. Images of the active region and flare from \textit{STEREO
  B} EUVI are shown in Figure~\ref{fig:euvi}.

The XRT on \textit{Hinode} is a high cadence, high spatial resolution (approximately
1\arcsec\ pixels) grazing incidence telescope that images the Sun in the soft X-ray and
extreme ultraviolet wavelength ranges. Temperature discrimination is achieved through the
use of focal plane filters. Because XRT can observe the Sun at short wavelengths, XRT
images can observe high temperature solar plasma very efficiently. The thinner XRT filters
allow longer wavelength EUV emission to be images and extend the XRT response to lower
temperatures. Further details on XRT are given in \cite{golub2007}.

\begin{figure*}
\centerline{%
  \includegraphics[clip,scale=1.00]{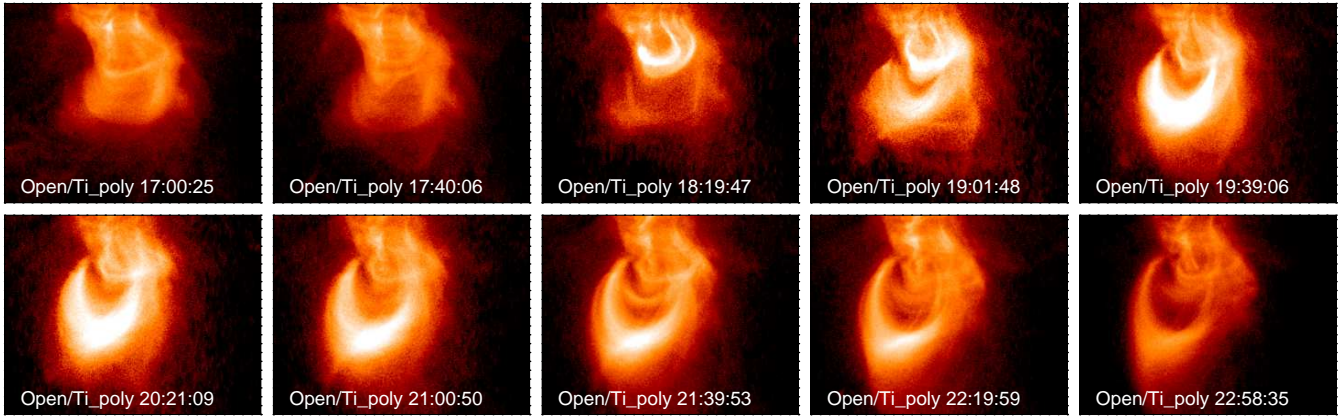}}
\caption{Selected XRT Open/Ti-Poly images from active region 10953 from 17 to 23 UT on May
  2, 2007. The field of view is $281\arcsec\times226\arcsec$.}
\label{fig:xrt}
\end{figure*}

As indicated in Figure~\ref{fig:summary}, the principal XRT images taken during this time
period were in three filters, Ti-Poly, Al-Thick, and Be-Thick, at a variable
cadence. Unfortunately, the exposure times on the Be-Thick images are too short for the
images to be used for analyzing active region loops. The standard processing routine
\verb+xrt_prep+ is used to remove the CCD bias, dark current, and calibrate the
images. Images are also ``dejittered'' using \textit{Hinode} spacecraft housekeeping data
so that the images are co-aligned with respect to each other.  \textit{Hinode} tracks
solar rotation so there is no need to account for it in the images. Example Ti-Poly images
are shown in Figure~\ref{fig:xrt}. These images have a field of view of
$512\arcsec\times512\arcsec$.

The \textit{TRACE} instrument is a high resolution normal incidence telescope. The primary
and secondary mirrors are divided into quadrants and a rotating shutter is used to select
which quadrant is illuminated. Three of the quadrants are coated with multilayers for
imaging at EUV wavelengths. The multilayer coatings have peak sensitivities at
approximately 171, 195, and 284\,\AA. The fourth quadrant is coated with aluminum and
magnesium fluoride for imaging very broad wavelength ranges near 1216, 1550, 1600, and
1700\,\AA. Images in all of the wavelengths are projected onto a single detector, a
$1024\times1024$ CCD. Each CCD pixel represents a solar area approximately 0.5\arcsec\ on
a side. The instrument is described in detail by \cite{handy1999}. The initial in-flight
performance is reviewed by \cite{golub1999} and \cite{schrijver1999}.

During this period \textit{TRACE} observed mainly in the 171\,\AA\ channel at a cadence of
about 60\,s with occasional 1600\,\AA\ and white light context images. All of the images
have a $512\arcsec\times512\arcsec$ field of view. As shown in Figure~\ref{fig:summary},
there are periodic data gaps in the TRACE data due to orbital eclipses. All of the
\textit{TRACE} images are processed using a standard application of
\verb+trace_prep+. Additionally, the images are despiked and co-aligned with respect to
each other to account for solar rotation and drifts in the pointing. A simple
cross-correlation method is used for this purpose. The \textit{TRACE} 171\,\AA\ images are
very similar in appearance to the EUVI 171\,\AA\ images.

\section{Analysis}

The primary goal of this study is to compare multi-thread, hydrodynamic simulations with
the emission observed in an evolving coronal loop. As we will discuss in more detail later
in the paper, hydrodynamic simulations involve solving the equations for the conservation
of mass, momentum, and energy in the loop given some input heating rate. Relating the
heating rate to physical observables is a critical element of the modeling. Our strategy
is to measure the electron density with EIS and use a family of hydrodynamic simulations
to infer the required heating rate for this density. Previous numerical simulations
suggest that for a fixed loop length there is a power law relationship between the peak
electron density and the input energy \citep{warren2004a}. Since we want the loop length
to be fixed, the other critical element of this modeling is an accurate measurement of the
loop geometry, including the inclination. Observations from the twin EUVI instruments
allow the loop geometry to be measured, and we use the \textit{STEREO} software package
developed for this purpose \citep{aschwanden2008b}.

Once the density and loop geometry have been determined we can perform hydrodynamic
simulations and synthesize the expected emission during the entire evolution of the
loop. The simulation results can then be compared with light curves determined from
\textit{TRACE} and XRT and in this section we discuss how these light curves are
calculated. The distribution of temperatures in the loops is an important constraint on
the modelings and in this section we also discuss the calculation of the differential
emission measure distribution with the EIS spectra.

\subsection{EIS}

Our analysis requires the identification of loops observed simultaneously with both EIS
and \textit{TRACE}. To facilitate this we wrote routines to co-align EIS rasters with
\textit{TRACE} images and to display 24 bit color images using an EIS raster for one color
channel and a TRACE image for another color channel. An animation of these images allowed
us to quickly identify times when the EIS slit was co-spatial with a loop observed with
\textit{TRACE}. \textit{TRACE}'s small field of view and frequent data gaps due to orbital
eclipse make finding good data sets more difficult than anticipated.  

To optimize the co-alignment between EIS and \textit{TRACE} we had to allow for a roll
angle between the images. This is in addition to the usual spatial shifts between the
pointing information contained in the data headers. Since the \textit{TRACE} data was
taken in the \ion{Fe}{9}/\textsc{x} 171\,\AA\ channel we used the EIS \ion{Fe}{10}
184.536\,\AA\ raster for co-alignment.

\begin{figure*}
\centerline{%
  \includegraphics[clip,scale=0.90]{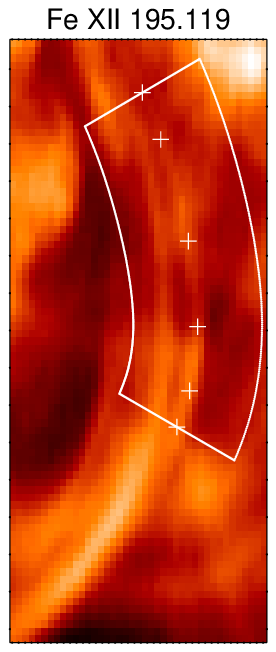}
  \includegraphics[clip,scale=0.90]{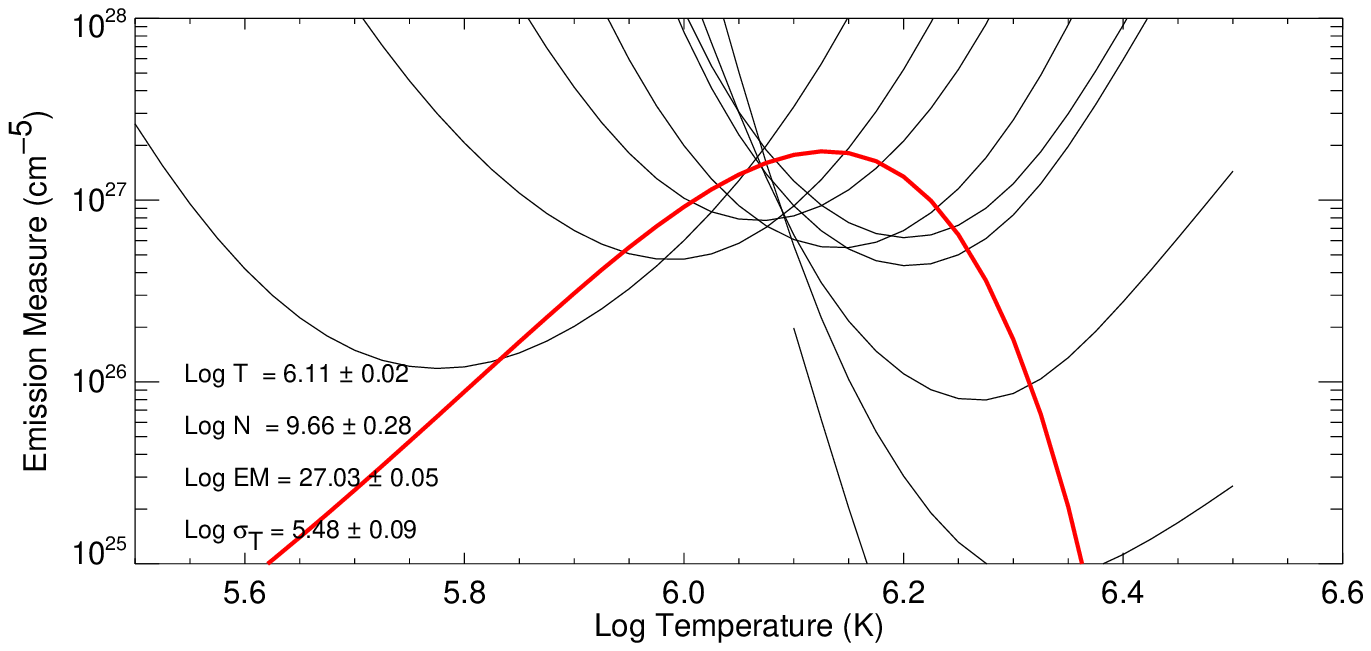}}
\centerline{%
  \includegraphics[clip,scale=0.90]{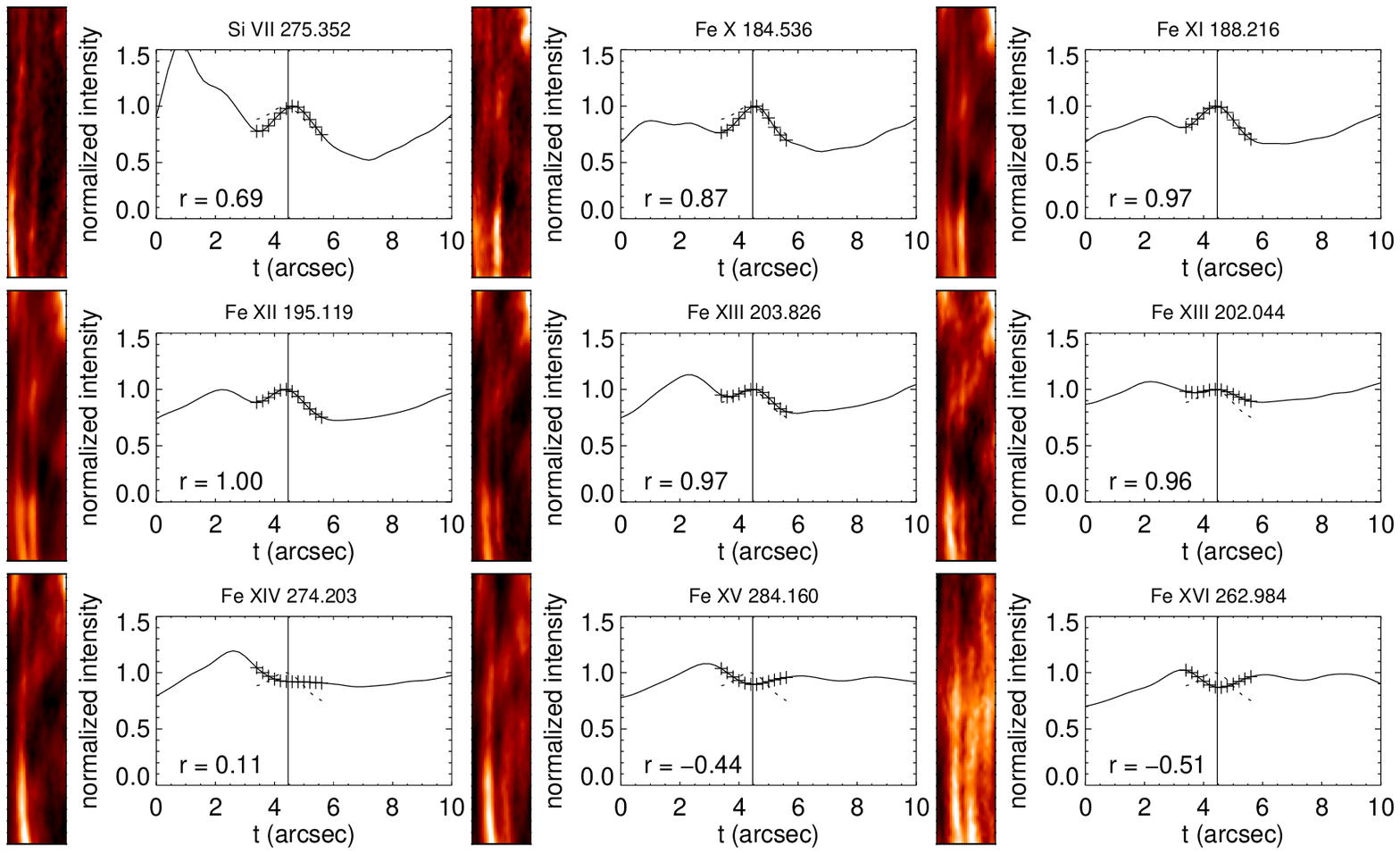}}
\caption{\small An EIS snapshot of an evolving loop. The top left panel shows the region
  surrounding the loop in \ion{Fe}{12} 195.119\,\AA. The pluses ($+$) are the spline knots
  used to define the coordinate system along the loop. The bottom panels show the
  straightened loop region in $s,t$ coordinates for 9 emission lines. The intensity
  averaged along the loop is also plotted for each line. The region used to compute the
  background subtracted loop intensities is indicated with the plus ($+$) symbols. For
  reference the \ion{Fe}{12} 195.119\,\AA\ intensities are also shown in each panel
  (dotted line) and the \ion{Fe}{12} 195.119\,\AA\ centroid is also shown in each panel
  (vertical line). Finally, the correlation between the \ion{Fe}{12} 195.119\,\AA\
  intensities and the loop intensities for each line are indicated. The loop emission is
  well correlated at temperatures below \ion{Fe}{13} and poorly correlated at higher
  temperatures (\ion{Fe}{14}--\textsc{xvi}). This temperature dependence is reflected in
  the differential emission measure shown in the top panel. Here the emission measure loci
  are shown (black lines) along with the DEM (red line).}
\label{fig:eis_ints}
\end{figure*}

The EIS intensities for the loop of interest are summarized in
Figure~\ref{fig:eis_ints}. Once the region of interest was identified we manually selected
spatial positions along the loop in the EIS \ion{Fe}{12} 195.119\,\AA\ raster. These
points are used as spline knots to define the loop coordinate system ($s,t$), with $s$
along the loop and $t$ perpendicular to it (see \citealt{aschwanden2008b} Figure 3). Since
these loop coordinates are not necessarily aligned to the CCD we have interpolated to
determine the intensities along the selected segment. The loop segments displayed in
Figure~\ref{fig:eis_ints} are interpolated to 0.2\arcsec\ per pixel. Using the loop
coordinate system it is a simple matter to compute the intensity averaged along the loop
segment. The coordinate system derived from \ion{Fe}{12} 195.119\,\AA\ is used for all of
the EIS rasters.

To further isolate the intensity of the loop we identify two background points and fit a
single Gaussian with a linear background to the selected region. Background subtraction is
essential to separating the intensity in the loop from the contribution of the ambient
corona, but there is no unique method for computing it. Analysis of EUVI data, which has
the advantage of providing two different lines of sight for a single loop, suggests that
the background subtracted intensities can be computed consistently, although there can be
considerable uncertainties for individual measurements \citep{aschwanden2008b}.

Since we are interested in emission that is co-spatial we extract the same region from all
of the EIS rasters. The resulting intensities are shown in Figure~\ref{fig:eis_ints}. To
test how co-spatial the emission is at various temperatures, we have calculated a simple
correlation coefficient between the background subtracted intensity in each line and
\ion{Fe}{12} 195.119\,\AA, which represents a middle ground between the highest and lowest
temperature emission that is observed.  For this loop well correlated, co-spatial emission
is observed for \ion{Fe}{13} and below. For the emission at higher temperatures
(\ion{Fe}{14}--\textsc{xvi}) the correlation is poor.

\begin{figure}
\centerline{%
   \includegraphics[clip,scale=0.475]{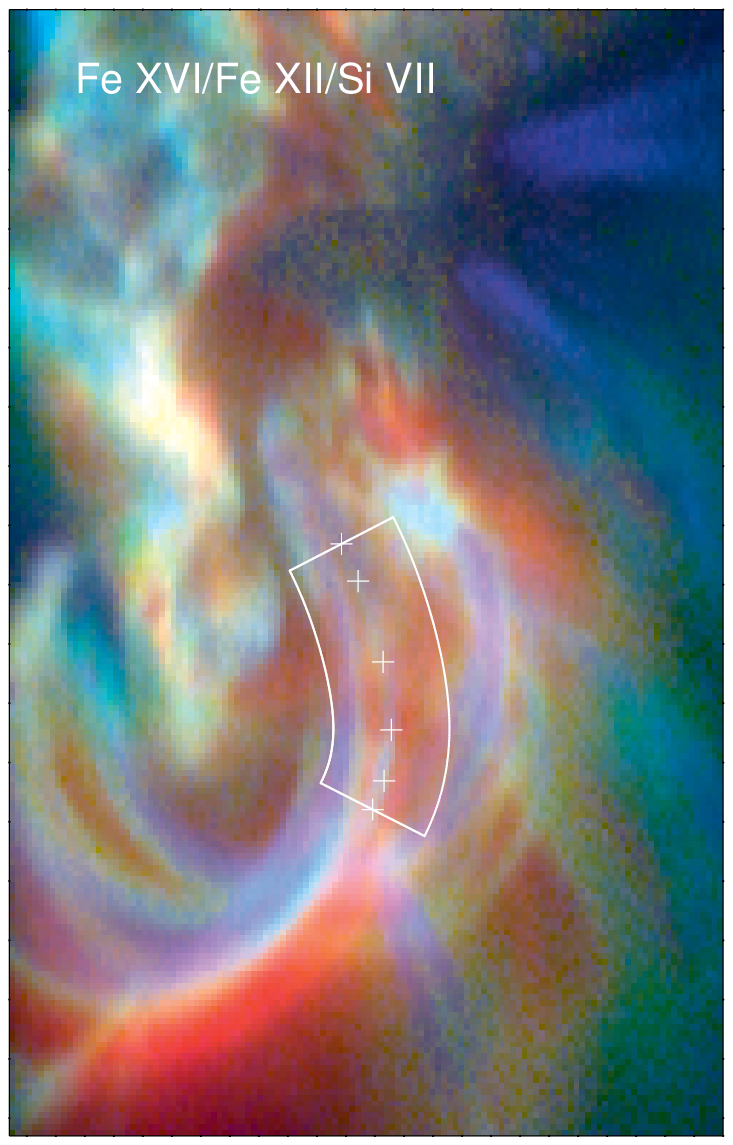}
   \includegraphics[clip,scale=0.475]{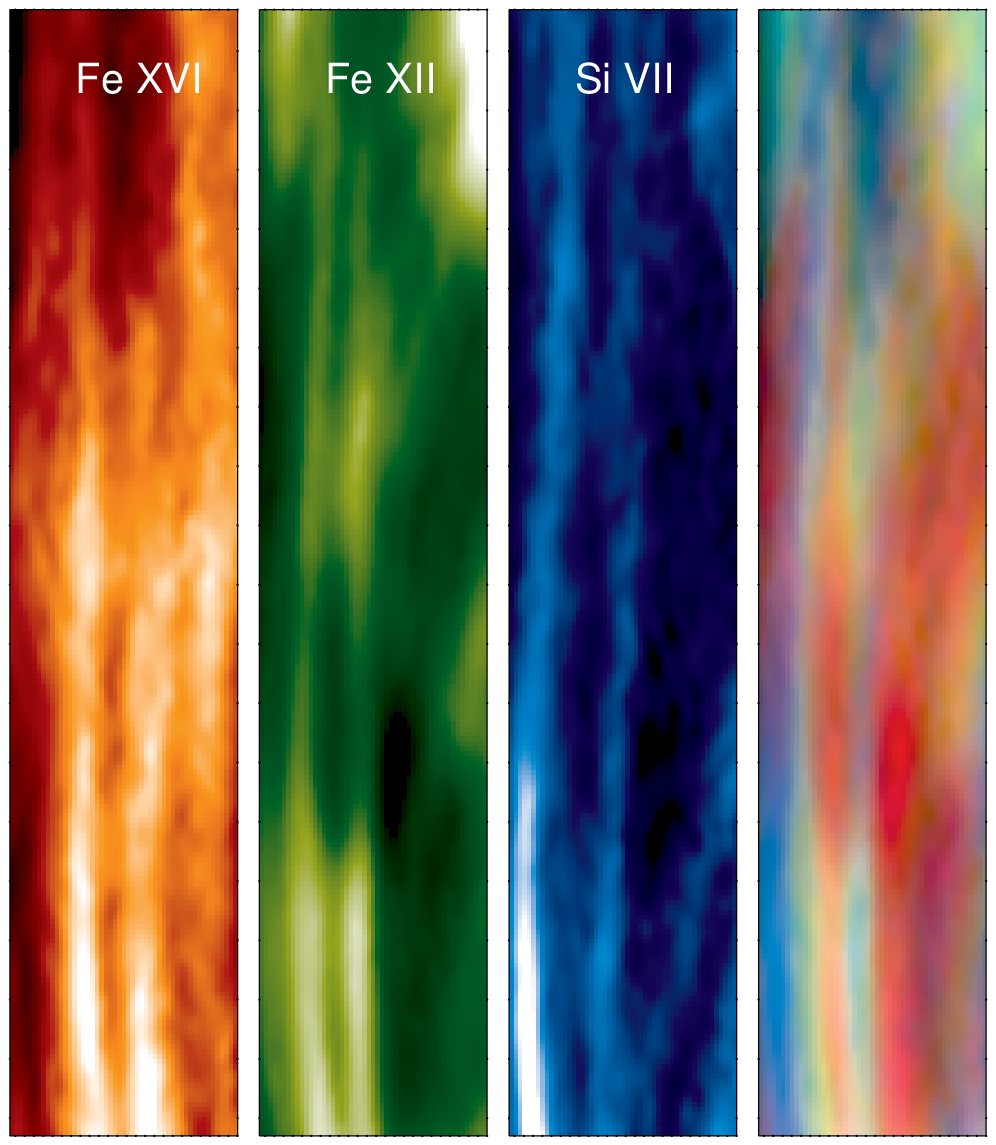}}
          \caption{EIS multicolor images composed of \ion{Fe}{16} 262.984\,\AA\ (red),
            \ion{Fe}{12} 195.119\,\AA\ (green), and \ion{Si}{7} 275.354\,\AA\ (blue). Each
            wavelength is used as one of the color channels in the composite 24 bit
            image. The panel on the left shows the active region, the panels on the right
            show the loop. The panel on the far right is the composite loop image. In
            regions where there is strong emission at multiple temperatures the image is
            white. The loop arcade is dominated by the primary colors suggested a
            relatively narrow distribution of temperatures the loops.}
\label{fig:eis_color}
\end{figure}

It is clear from the images shown in Figures~\ref{fig:eis} and \ref{fig:eis_ints} that the
combination of high spatial resolution and good temperature discrimination allows EIS to
probe the interrelationship of emission at different temperatures. The loop intensities
suggest that the emission at different temperatures is generally not co-spatial and that
the DEM in the loop of interest should be relatively narrow. This is shown more clearly in
the ``multicolor'' image of this region presented in Figure~\ref{fig:eis_color}. This
image, which is a 24 bit image formed from rasters in 3 different emission lines, would be
white in regions where the DEM is broad and the emission is strong in all three
lines. There are some composite regions that are cyan (green $+$ blue), but the post-flare
loop arcade is generally dominated by emission in the primary colors suggesting relatively
narrow distributions of temperature in each loop.

To investigate the temperature structure of this loop more quantitatively we compute the
differential emission measure using the background subtracted loop intensities for the
loop segment. The intensities are related to the differential emission measure by the
usual expression
\begin{equation}
I_\lambda = \frac{1}{4\pi}\int\epsilon_\lambda(n_e,T)\xi(T)\,dT,
\label{eq:ints}
\end{equation}
where $\epsilon_\lambda(n_e,T)$ is the plasma emissivity and $\xi(T)$ is the differential
emission measure distribution.  We consider a Gaussian DEM model
\begin{equation}
\xi(T) = \frac{EM_0}{\sigma_T\sqrt{2\pi}}
    \exp\left[-\frac{(T-T_0)^2}{2\sigma_T^2}\right],
\end{equation}
which allows for a dispersion in the temperature distribution. Since the density is an
important parameter in determining the emissivities of many of these lines we leave it is
a free parameter. To determine the best-fit parameters ($EM_0$, $T_0$, $\sigma_T$, $n_e$)
we use a Levenberg-Marquardt technique implemented in the \verb+MPFIT+ package. The
CHIANTI 5.2 atomic physics database (e.g., \citealt{landi2006}) is used to calculate the
emissivities. The abundances of \cite{feldman1992} and the low density ionization
fractions of \cite{mazzotta1998} are assumed.

There are several subtleties to computing the emission measure parameters. One is the
statistical uncertainty associated with each intensity. Because the intensities are
averaged over a significant area the statistical uncertainties are generally small. The
systematic errors introduced by the background subtraction and the atomic data, however,
are large, but difficult to estimate. We simply assume that the relative error is 20\% of
the measured intensity. Another question is how to deal with the emission from lines such
as \ion{Fe}{16} 262.984\,\AA, that do not show evidence for the loop and have no measured
intensity. For these lines the background subtracted intensity is zero. Observations of
these lines provide important constraints on the high temperature component of the DEM and
must be included. To account for possible errors in computing the background subtracted
intensities for these lines we assume that the uncertainty in the intensity is 20\% of the
background instead of 20\% of the loop intensity. Since the background can be large this
represents a substantial enhancement of the uncertainty.

The resulting DEM for this loop segment is shown in Figure~\ref{fig:eis_ints}. The
observed and computed intensities are given in Table~\ref{table:ints}. From this analysis
we obtain an electron density of $\log n_e=9.7$. For this work we use the \ion{Fe}{13}
203.826/202.044\,\AA\ lines to provide the bulk of the density sensitivity. Recently
\cite{young2009b} have noted systematic discrepancies between the various density
sensitive line ratios from \ion{Fe}{12} and \ion{Fe}{13}. In light of this we compared the
densities inferred from \ion{Fe}{13} 203.826/202.044\,\AA\ and \ion{Fe}{12}
186.880/195.119\,\AA\ with those from \ion{Si}{10} 258.375/261.058\,\AA\ in a series of
other active region and quiet Sun observations. The \ion{Si}{10} lines, which were not
included in this study, are relatively weak and not sensitive over as large a range as the
\ion{Fe}{12} and \ion{Fe}{13} lines. However, the atomic data for \ion{Si}{10} is
potentially more reliable than the atomic data available for the complex Fe ions. We find
that the densities derived from the \ion{Si}{10} and \ion{Fe}{13} ratios are in excellent
agreement and emphasize the \ion{Fe}{13} ratio here. The densities inferred from
\ion{Fe}{12} 186.880/195.119\,\AA\ are as much as a factor of 3 higher. Details of this
analysis will be presented in a future paper.

The dispersion in the temperature is found to be $\log\sigma_T=5.5$, which is comparable
to other active region loop observations with EIS \citep{warren2008b}.  These lines
observed in the quiet corona above the limb indicate much narrower temperature
distributions ($\log\sigma_T\lesssim5.0$, \citealt{warren2009a}). Here we find a dispersion
in temperature that is several times greater, indicating that this loop is not strictly
isothermal. The relatively intense, co-spatial emission observed from both \ion{Si}{7} and
\ion{Fe}{13} provides the best direct evidence for a distribution of temperatures in this
loop. These lines have peak temperatures of formation that are about 1\,MK apart. The
application of a delta function emission measure to Equation~\ref{eq:ints} confirms that a
single temperature model cannot adequately reproduce the intensities observed in these
emission lines.

\subsection{TRACE}

In searching for coronal loops observed with both EIS and \textit{TRACE} we co-aligned the
EIS rasters with the \textit{TRACE} images for this period. Thus to compute the
\textit{TRACE} intensities we use the same coordinate system and apply it to all of the
\textit{TRACE} images taken during the time of interest. Since the co-alignment between
EIS and \textit{TRACE} is not perfect, the spline knots selected in the EIS raster are
modified slightly to better align with the loop. Background subtracted intensities are
computed for each of the available \textit{TRACE} images taken during this time using the
same procedure that was used on the EIS data. For each image the region defined by the
spline knots was extracted, straightened, and averaged along the loop coordinate to
produce the intensity as a function of the perpendicular coordinate. The intensity at each
time refers to the background subtracted intensity integrated over the loop.

\begin{figure*}[h!]
\centerline{%
  \includegraphics[clip,scale=0.65]{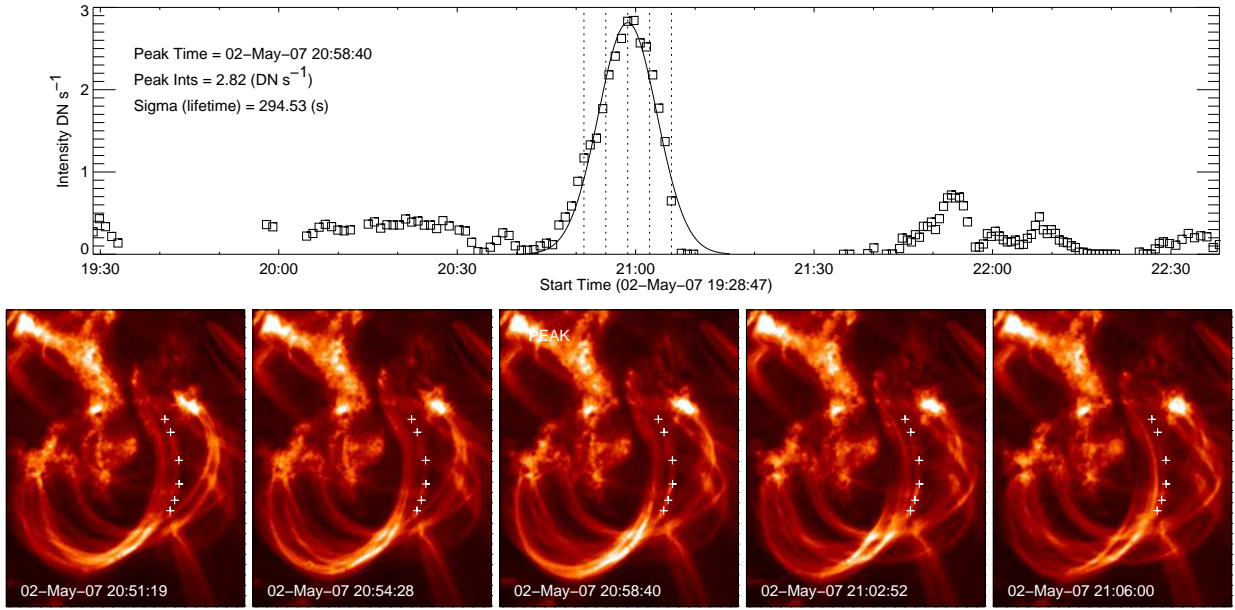}}
\caption{TRACE 171\,\AA\ intensities as a function of time for the loop observed with
  EIS. The TRACE and EIS images near the peak of the event have been co-aligned. The
  vertical lines indicate the times for the TRACE images shown in the figure.}
\label{fig:trace_ints}
\end{figure*}

The \textit{TRACE} light curve is shown in Figure~\ref{fig:trace_ints}. For most of this
period the background subtracted loop intensities are in the noise and the evolution of
this loop can be seen clearly. We have manually selected the region around the peak in the
light curve and fit it with a single Gaussian. This fitting yields a Gaussian width of
about 294\,s.

\subsection{EUVI}

To extract the three dimensional geometry of this loop we use the software package
developed by Markus Aschwanden and the \textit{STEREO} team for this purpose. An
application of this software is discussed in detail in \cite{aschwanden2008a} and
\cite{aschwanden2008b}.  The initial processing co-registers images from \textit{STEREO A}
and \textit{B} to account for differences in spacecraft roll angle and spatial
resolution. The next step is to outline the loop in the \textit{STEREO A} image. The
selected coordinates are projected onto the corresponding \textit{B} image. Since geometry
of the loop has not yet been determined this projection yields a range of possible
coordinates in the \textit{B} image and the user selects the position of the loop within
this range. Once the coordinates of the loop have been selected in both images the
geometry of the loop is determined from simple trigonometric relationships.

\begin{figure*}
\centerline{%
  \includegraphics[clip,scale=1.0]{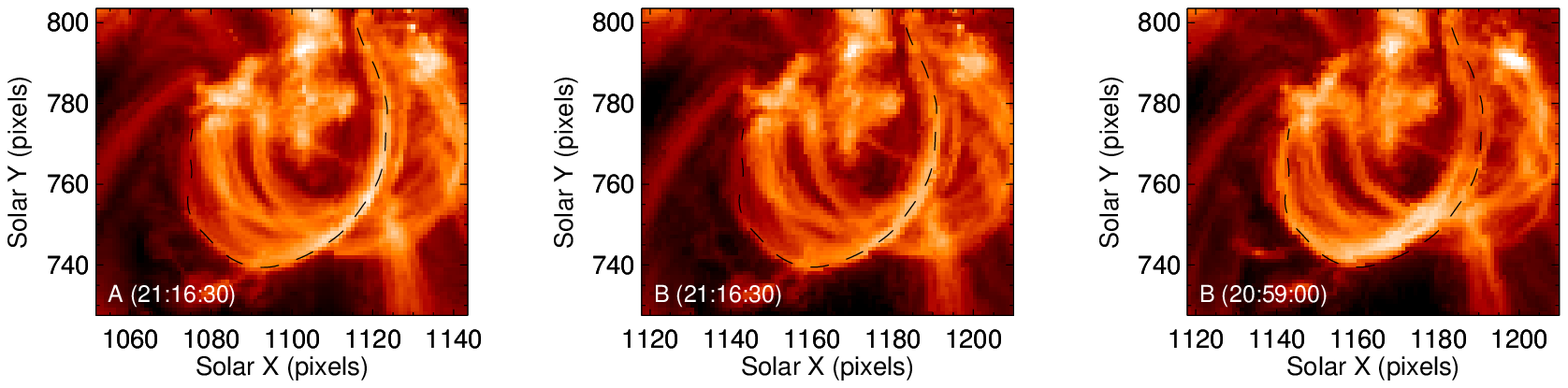}}
\centerline{%
  \includegraphics[clip,scale=0.95]{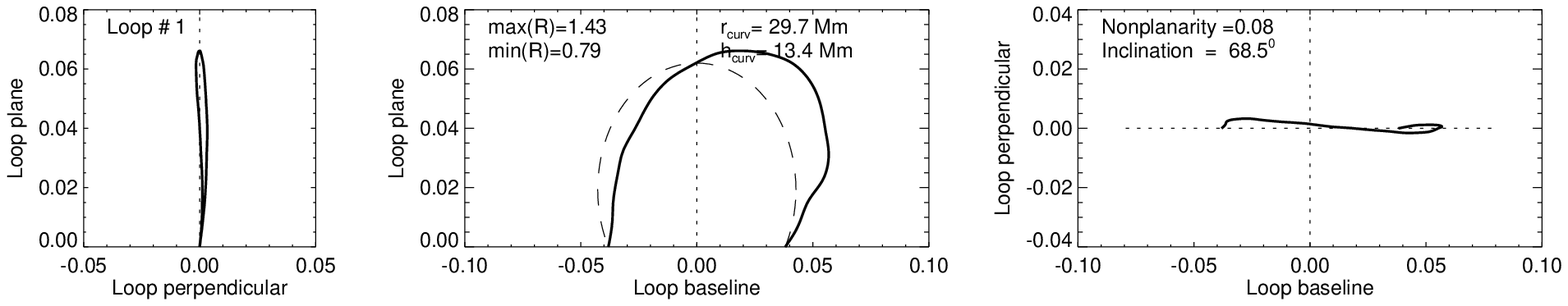}}
\caption{\textit{STEREO} EUVI reconstruction of the loop geometry. Because of a data gap
  with \textit{STEREO B}, images taken May 2, 2007 21:16 UT, slightly after the loop
  appears in \textit{TRACE}, are used. The top panels show the loop traced out in both
  images. The image on the far left is the \textit{STEREO B} 171\,\AA\ image taken near
  the peak in the \textit{TRACE} 171\,\AA\ light curve. The bottom panels show the loop in
  various planes. The loop length is 135\,Mm and the loop inclination is 68.5$^\circ$.}
\label{fig:euvi_stereo}
\end{figure*}

Due to the gap in the \textit{A} data, an image pair at 21:16 UT, after the peak emission
observed in this loop, is used, so there is some ambiguity in the loop identification. The
selection of nearby structures in the loop arcade generally yield very similar
results. The projection of the extracted loop geometry onto the EUVI \textit{B} image
available at the peak of the loop emission (20:59 UT) also outlines the observed loop very
well. Thus it does not appear that the calculation of the loop geometry is significantly
impacted by the data gap. The selected loop and the projection of this loop in various
planes is shown in Figure~\ref{fig:euvi_stereo}.

\subsection{XRT}

\begin{figure*}
\centerline{%
 \includegraphics[clip,scale=0.65]{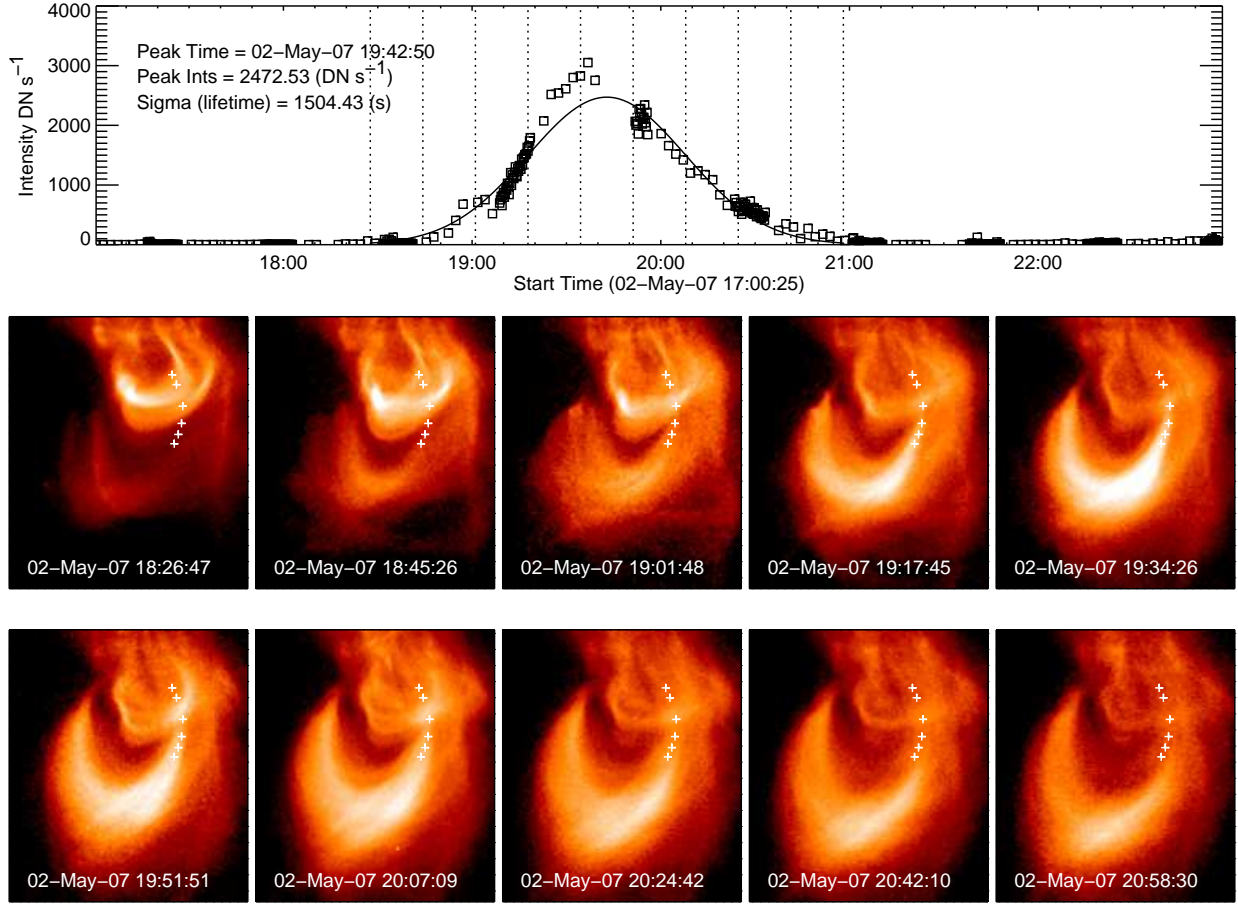}}
\caption{XRT Ti Poly intensities as a function of time for the region observed with
  EIS. The vertical lines indicate the times for the XRT images shown in the figure. The
  crosses indicate positions along the loop used to define the loop coordinate system.}
\label{fig:xrt_ints}
\end{figure*}

Images taken with the XRT provide information on the evolution of the loop at high
temperatures. We have computed an XRT light curve similar to that computed for
\textit{TRACE}. To use the loop coordinates derived from EIS we first co-align the XRT
images with the EIS \ion{Fe}{16} 262.984\,\AA\ raster, which shares some common
features. We then compute the background subtracted intensities by selecting two
background points and doing a linear fit. The emission seen earlier in the event is much
broader than what is observed TRACE and so we select a wider area to compute the
intensities. The resulting XRT light curve is shown in Figure~\ref{fig:xrt_ints}.

\begin{figure*}
\centerline{%
  \includegraphics[clip,scale=0.675]{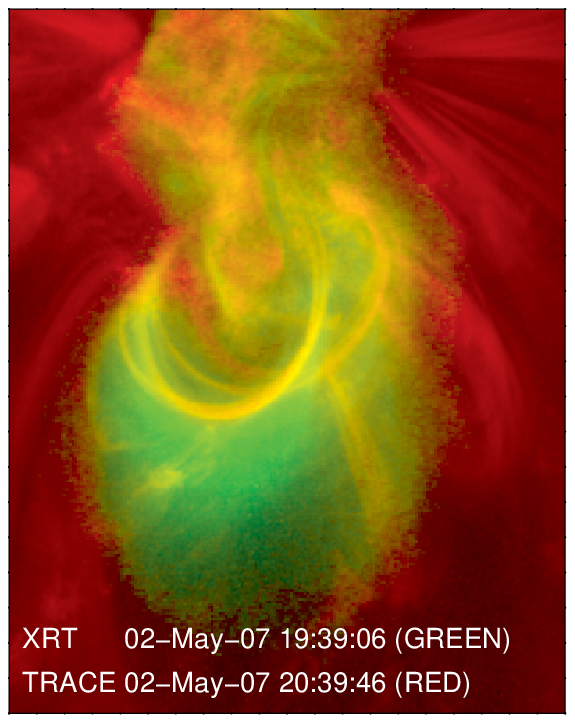}
  \includegraphics[clip,scale=0.675]{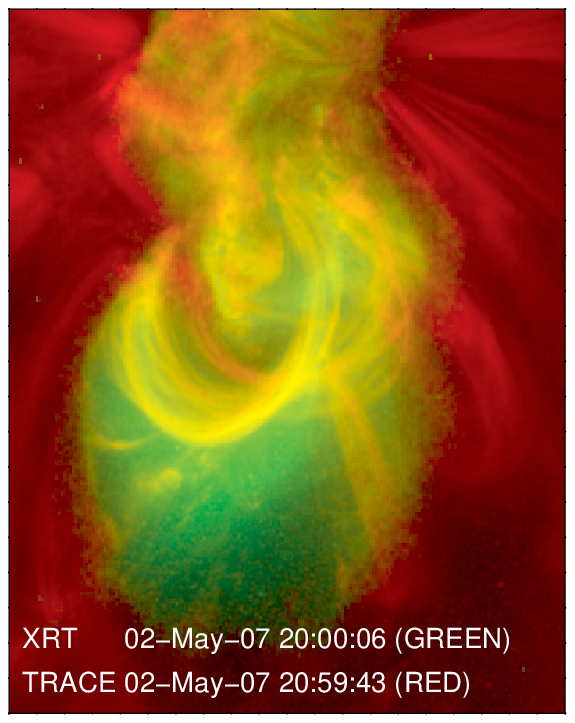}
  \includegraphics[clip,scale=0.675]{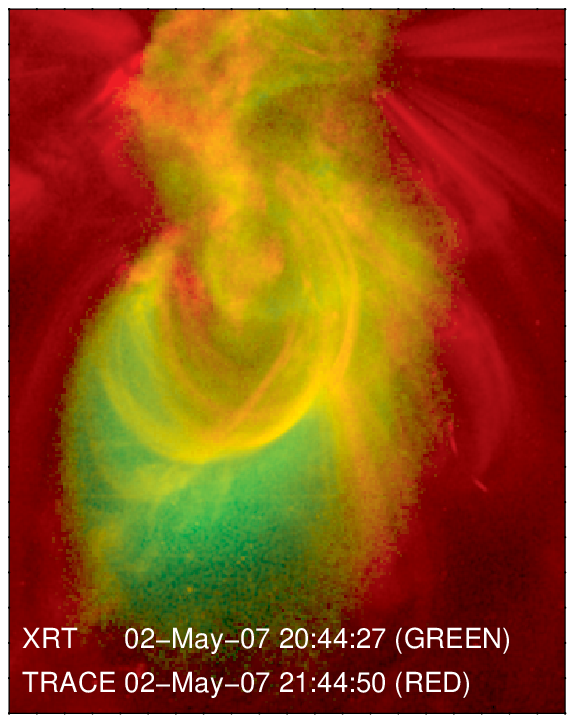}
  \includegraphics[clip,scale=0.675]{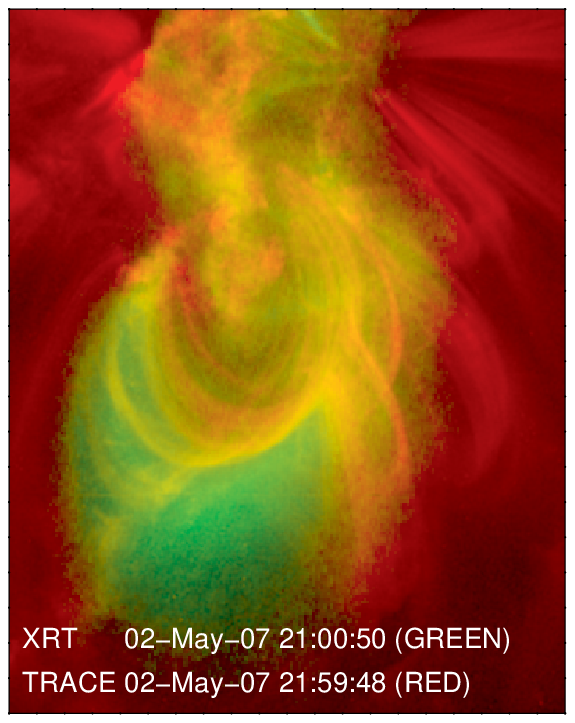}
}
\caption{Multicolor images from XRT and \textit{TRACE}. The red channel is \textit{TRACE}
  and the green channel is XRT. The image times are offset by one hour to illustrate the
  differences in morphology of post-flare loops in different stages of evolution. The high
  temperature XRT images do not show the fine loops observed at lower temperatures in
  \textit{TRACE}.}
\label{fig:xrt_trace}
\end{figure*}

\begin{figure*}
\centerline{%
 \includegraphics[clip,scale=1.0]{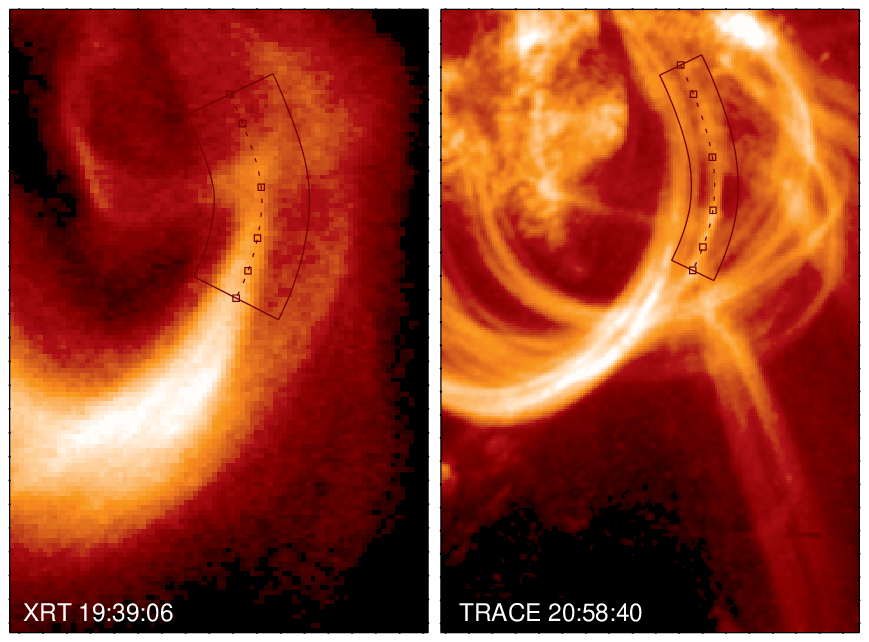}
 \includegraphics[clip,scale=1.0]{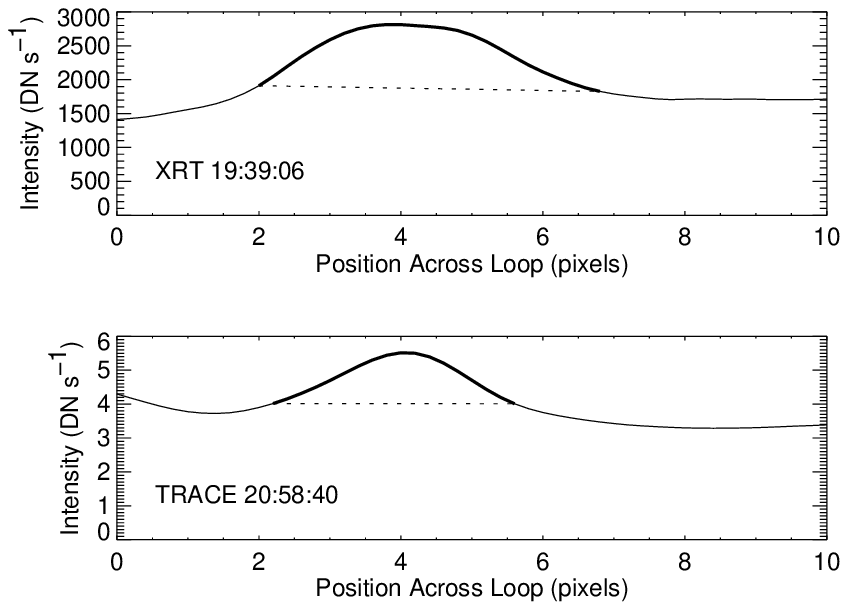}}
\caption{Left panels: XRT and TRACE images of the loop at different times and
  temperatures. The region used to compute the loop intensities are indicated. Right
  panels: The average XRT and TRACE intensities across the loop. The dotted line indicates
  the assumed background.}
  \label{fig:xrt_trace_ints}
\end{figure*}

As is indicated by the light curve, the XRT images clearly show strong emission in the
region that is eventually occupied by the loops observed with EIS and \textit{TRACE}. It
is also clear, however, that XRT does not show any individual loops that are as narrow as
those that are seen at cooler temperatures. Consistent with this, the loop cross section
measured with XRT is systematically wider than what is measured with EIS and
\textit{TRACE}. To illustrate these differences we have constructed multicolor images from
XRT and \textit{TRACE}. These images, which are presented in Figure~\ref{fig:xrt_trace},
use different color channels to display the XRT and \textit{TRACE} data in the same
picture. To illustrate the differences in morphology as the plasma cools we have offset
the times of the selected images by one hour. We show examples of the loop cross sections
in Figure~\ref{fig:xrt_trace_ints}. This difference between high temperature and low
temperature emission in flares and active regions is well documented
\citep[e.g.,][]{tripathi2009,patsourakos2002,warren2000b}, and may be further evidence for
fine scale structure in the solar corona. We will discuss this point in more detail later
in the paper.

\section{Hydrodynamic Modeling}\label{sec:modeling}

One of the primary paradoxes of coronal loops with temperatures near 1\,MK is the
disparity between the rapid cooling suggested by the high electron densities and the
relatively long observed lifetimes. EIS density diagnostics allow us to make more rigorous
comparisons between these time scales. Densities inferred from observed intensities
requires an accurate determination of the differential emission measure, the absolute
instrumental calibration, and the loop geometry. This measurement also represents a lower
bound on the density. The density inferred from density sensitive line ratios circumvents
many of these problems.

If we assume that the loop is cooling only through radiation the energy balance is simply
\begin{equation}
\frac{\partial E}{\partial t} \approx -n_e^2\Lambda(T_e),
\end{equation}
where $\Lambda(T_e)$ is the radiative loss function for an optically thin plasma. In the
limit of no flows the energy is $E = \frac{3}{2}P$. The plasma pressure given by $P =
2n_ek_BT_e$, with $k_B$ is the Boltzmann constant.  The radiative cooling time is 
defined as
\begin{equation}
\frac{1}{\tau_R} = \frac{1}{E}\frac{\partial E}{\partial t} = 
                   \frac{1}{P}\frac{\partial P}{\partial t},
\end{equation}
and is given by
\begin{equation}
\tau_R = \frac{3k_BT_e}{n_e\Lambda(T_e)}.
\end{equation}
Using the temperature and density derived from the EIS DEM analysis ($\log n_e\simeq 9.7$
and $\log T_e\simeq 6.11$) and a radiative loss rate of $3.2\times10^{-22}$\,erg cm$^3$
s$^{-1}$ \citep{brooks2006} we obtain a radiative cooling time of 341\,s.

The radiative cooling time is not directly comparable to the loop lifetime that we have
measured with \textit{TRACE}. If we make the additional assumptions that $T_e(t) =
T_0\exp(-t/\tau_T)$ and $T_e(t)\sim n_e^\alpha(t)$ we can relate the radiative
cooling time ($\tau_R$) to the timescale for changes in the temperature ($\tau_T$)
\begin{equation}
\tau_T = \frac{1+\alpha}{\alpha}\tau_R.
\end{equation}
Numerical simulations suggest that $\alpha\approx2$ \citep{jakimiec1992}. Finally, to
compare with the observed loop lifetime we must incorporate the temperature changes into
the TRACE temperature response curve. This yields
\begin{equation}
I_\lambda(t) \sim \exp\left[-\frac{T_\lambda^2}{\sigma_\lambda^2}
       \frac{(t-t_0)^2}{2\tau_T^2}\right]
\end{equation}
where $T_\lambda$ is the peak temperature of the TRACE response and $\sigma_\lambda$ is
the Gaussian width of the TRACE response (see \citealt{warren2003} Equation 9). This
yields
\begin{equation}
\sigma_t = \frac{\sigma_\lambda}{T_\lambda}\tau_T = 
           \frac{\sigma_\lambda}{T_\lambda}\frac{1+\alpha}{\alpha}\tau_R.
\end{equation}
With $T_{171} = 0.96$\,MK and $\sigma_{171} = 0.25$\,MK \citep{warren2003} we obtain
$\sigma_t = 141$\,s, which is smaller than the observed Gaussian width of 294\,s.

This mismatch between the predicted and observed loop lifetime is one of the key
motivations for the multi-thread modeling of coronal loops. By assuming that the observed
emission comes from a series of loops that are heated at different times it is clear that
we can create a composite loop with the required lifetime. The challenge is to also match
the relatively narrow DEM and the lifetime of the loop as it is observed with XRT.

To simulate the evolution of this loop we consider numerical solutions to the full
hydrodynamic loop equations using the NRL Solar Flux Tube Model (SOLFTM). We adopt many of
the same parameters and assumptions that were used in previous simulations with this code
and we refer the reader to the earlier papers for additional details on the numerical
model (e.g., \citealt{mariska1987,mariska1989}). For example, we assume that the loop is
symmetric and only simulate the evolution over half of the loop length. We also assume
that each loop has a constant cross section.

For this work we consider a heating function for each thread that is a simple
spatially uniform heating rate
\begin{equation}
  E_i(t) = E_0 + g(t)E_F^i,
\end{equation}
where $g(t)$ is a triangular envelope that peaks at time $t_i$ and has width $\delta$,
which we set to be 100\,s. The parameter $E_0$ is a small background heating rate that
provides the initial loop equilibrium. As we mentioned earlier, our strategy for inferring
the heating rates from the observations is to use the density determined from EIS and the
results from systematic hydrodynamic simulations. This will determine the peak heating
rate for the ensemble of threads. The heating rate for the other threads will be
determined by assuming a Gaussian envelope for the heating function.

\begin{figure}[t!]
\centerline{%
  \includegraphics[clip,scale=0.675]{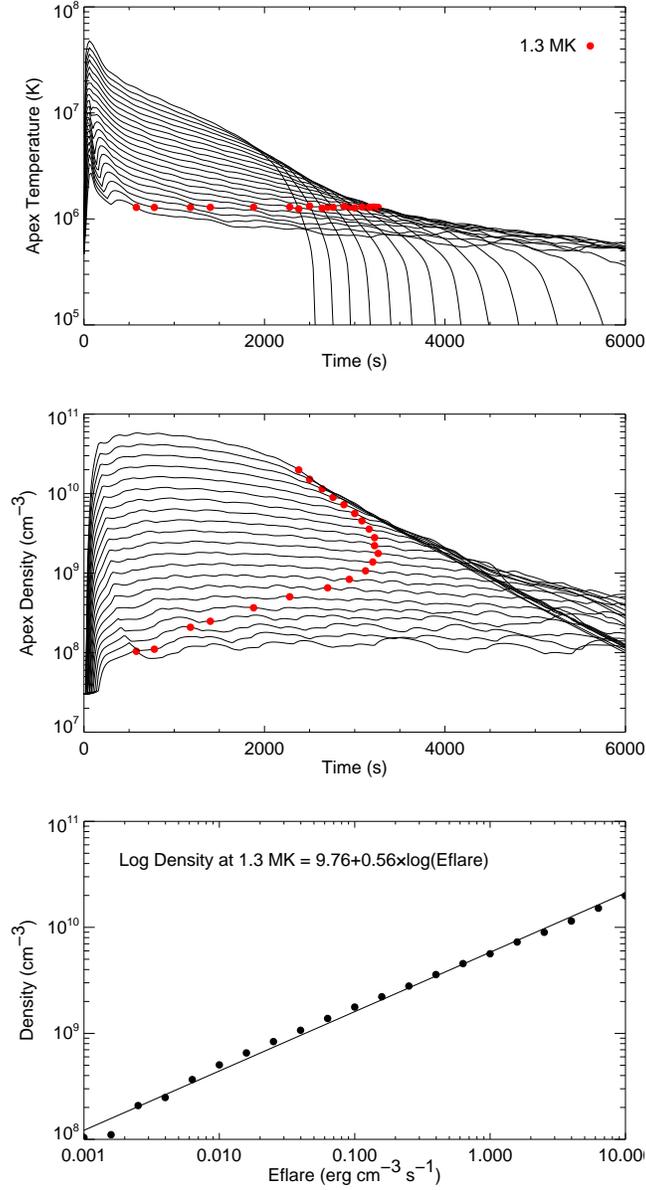}}
\caption{A family of hydrodynamic simulations for the loop geometry inferred from
  \textit{STEREO}. The top panels show the apex densities and temperatures as a function
  of time. For each simulation we have determined the density as the loop cools through
  1.3\,MK. The bottom panel shows the relationship between the input heating rate and the
  density at 1.3\,MK. }
\label{fig:apex}
\end{figure}

To determine the relationship between the input heating rate and the densities and
temperatures observed during the cooling phase of the loop evolution, we have performed 21
simulations with $E_F$ varying between $10^{-3}$ and $10^2$\,erg~cm$^{-3}$~s$^{-1}$. Using
the parameters derived from the EUVI observations, the loop length is fixed at 135\,Mm and
the loop inclination is fixed at 68.5$^\circ$. For each simulation we average over the
loop apex to determine a representative density and temperature. The simulation results
are summarized in Figure~\ref{fig:apex}. We find that the apex density at 1.3\,MK, which
is the peak temperature in the DEM, essentially scales as $n_e\sim\sqrt{E_F}$. This
implies that for a fixed loop length the observed intensity is linearly proportional to
the input energy. The relationship shown in Figure~\ref{fig:apex} indicates that a heating
rate of 0.8\,erg~cm$^{-3}$~s$^{-1}$ is required to reproduce an apex density of $\log n_e
\simeq 9.7$.

\begin{figure}[t!]
\centerline{%
  \includegraphics[clip,scale=0.675]{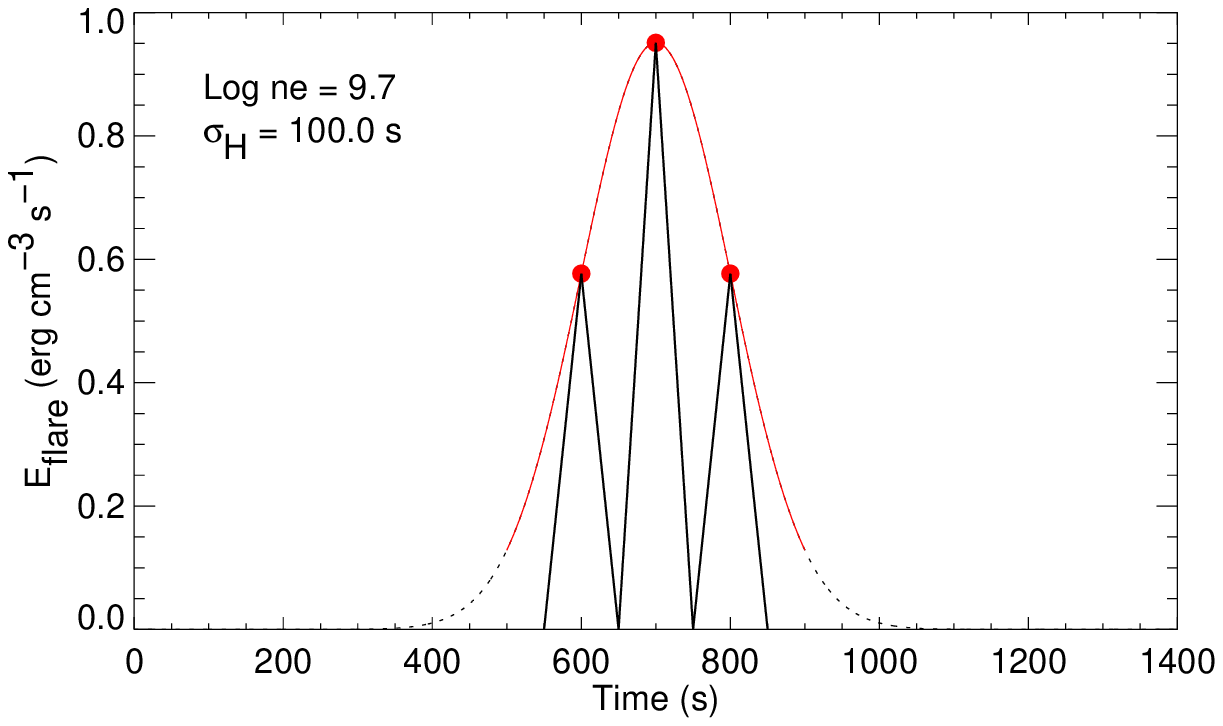}}
\centerline{%
  \includegraphics[clip,scale=0.675]{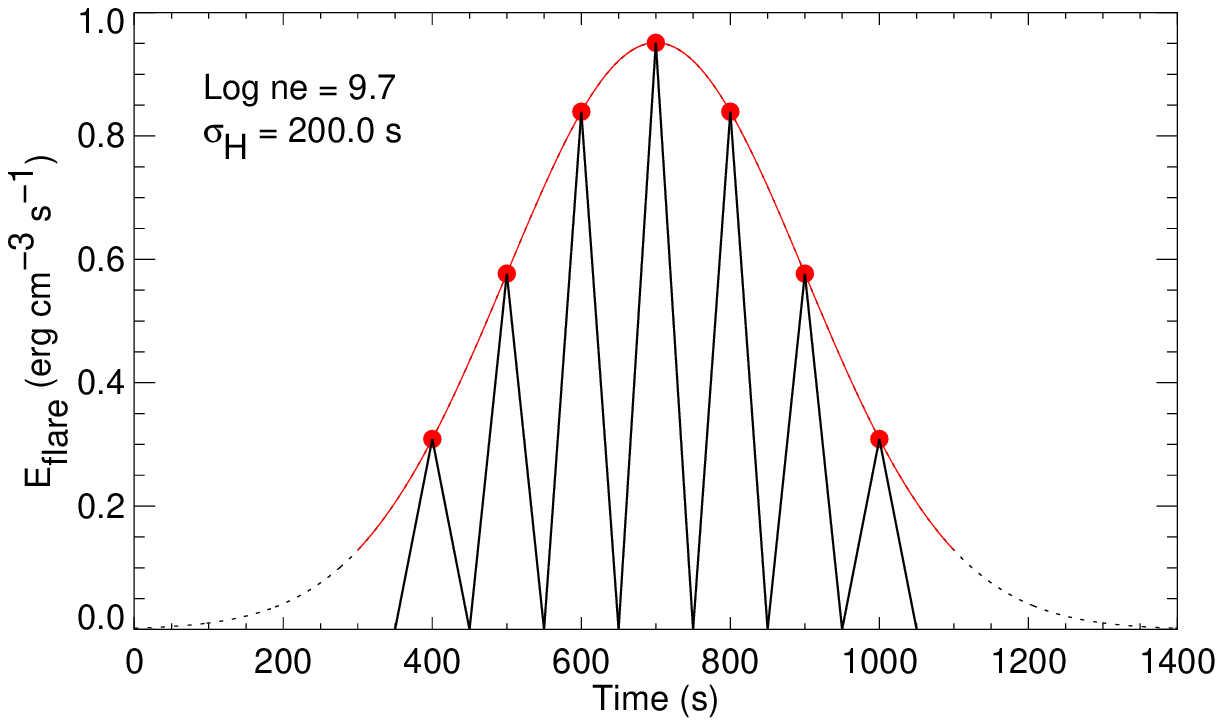}}
\centerline{%
  \includegraphics[clip,scale=0.675]{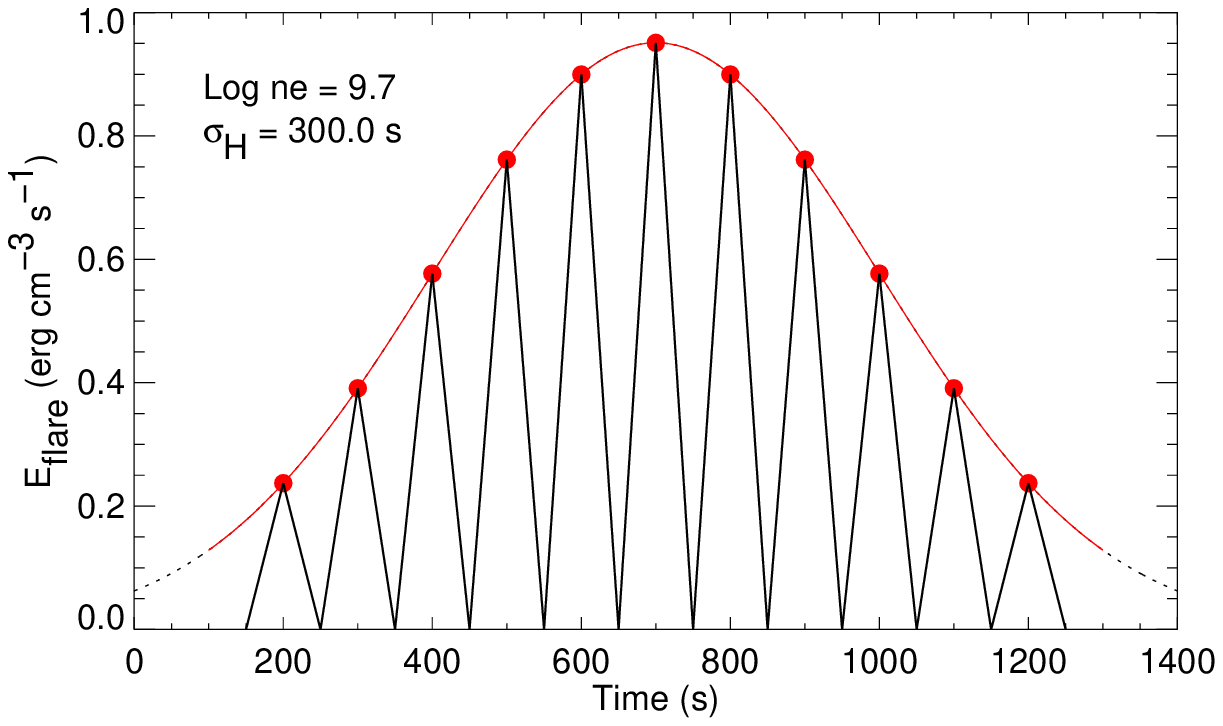}}
\caption{The heating rate for multi-thread simulations of the loop. The Gaussian curve
  represents the heating rate envelope. The triangles represent the heating rate for each
  thread.}
\label{fig:eflare}
\end{figure}

The density-heating rate relationship is valid for a single loop. For a multi-thread
simulation we assume the heating rate for each thread is related to this peak heating rate
by
\begin{equation}
E_F^i = E_F^{peak}\exp\left[-\frac{(t_i-t_0)^2}{2\sigma_H^2}\right],
\end{equation}
where $\sigma_H$ determines the duration of the heating envelope. The parameter $t_0$ is
chosen so that all of the times are positive. The heating events are spaced so that as the
heating in the previous loop ends the heating in the next loop begins. This heating
scenario is illustrated in Figure~\ref{fig:eflare} for $\sigma_H = 100$, 200, and 300\,s.

\begin{figure}[t!]
\centerline{%
  \includegraphics[clip,scale=0.675]{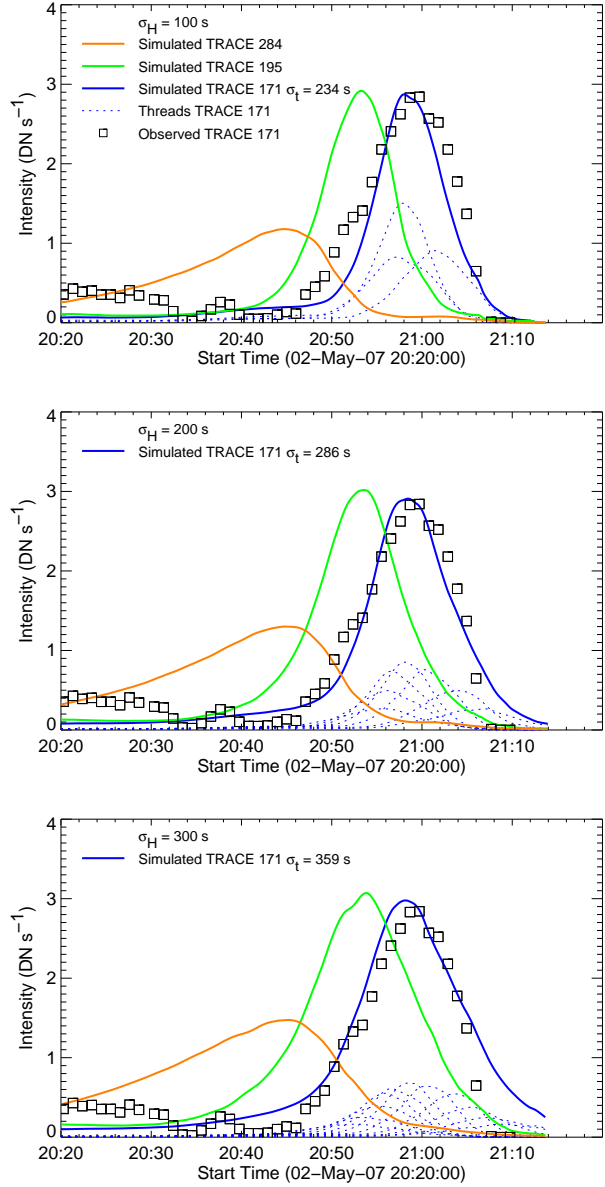}}
\caption{Simulated and observed TRACE light curves for $\sigma_H = 100$, 200, and
  300\,s. The simulation times have been adjusted so that the peak emission in 171\,\AA\
  matches what is observed. The magnitude of the simulated emission has also been scaled
  to match the observations. The lifetime of the loop is calculated from a Gaussian fit to
  the composite light curve. The $\sigma_H = 200$\,s simulation most closely matches to
  observed loop lifetime.}
\label{fig:trace_sim}
\end{figure}

Once the individual hydrodynamic simulations are run, we average over the loop apex at
each time step to compute a representative temperature and density. These densities and
temperatures are then used as inputs to the \textit{TRACE} temperature response to
calculate the expected count rates in the \textit{TRACE} 171, 195, and 284\,\AA\ channels
as function of time. The simulation times are shifted so that the peak in the 171\,\AA\
emission corresponds to the observed peak. Since we are not interested in resolving the
differences in absolute calibration among the various instruments, we also introduce a
scaling factor so that the peak simulated emission matches what is observed. The
simulation results are shown in Figure~\ref{fig:trace_sim} and indicate that
$\sigma_H=200$\,s simulation, which yields a simulated loop lifetime of $\sigma_t =
286$\,s, best matches the observations.

\begin{figure}[t!]
\centerline{%
  \includegraphics[clip,scale=0.675]{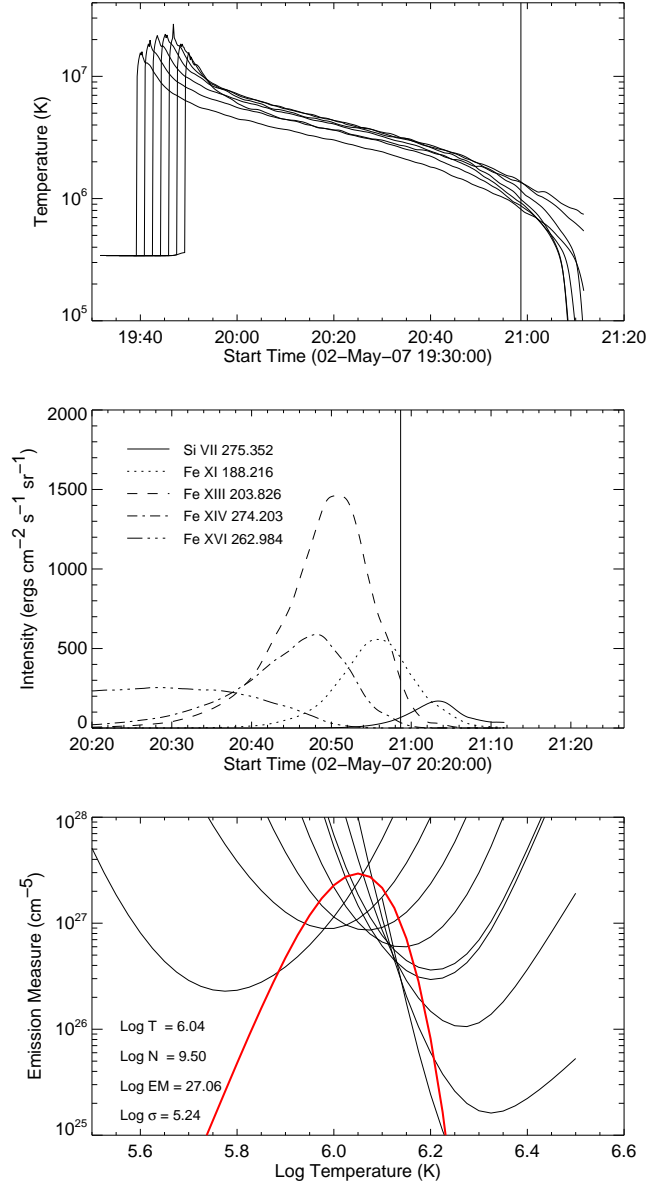}}
\caption{(\textit{top panel}) The apex temperature as a function of time for each
  thread. The vertical line represents the approximate time of the EIS
  observations. (\textit{middle panel}) Simulated EIS light curves for selected emission
  lines. Only the composite light curves are shown. (\textit{bottom panel}) The DEM
  inferred from the simulated EIS emission.}
\label{fig:eis_sim}
\end{figure}

We also use the simulated densities and temperatures as a function of time to compute the
expected intensities in many of the emission lines that can be observed with EIS. Light
curves from selected emission lines are shown in Figure~\ref{fig:eis_sim}. Since the
absolute time for the simulation has been established through the comparisons with
\textit{TRACE}, we select the simulated EIS intensities that correspond to the time of the
EIS observations and use them as inputs to the same differential emission measure code
that was used to produce Figure~\ref{fig:eis_ints}. The resulting simulated DEM is shown in
Figure~\ref{fig:eis_sim}. The simulated intensities are given in Table~\ref{table:ints}.

The agreement between the observed and simulated differential emission measure is
relatively good. The simulation captures the salient features of the observations, a
relatively high density and a narrow DEM. For these simulated intensities we do obtain a
somewhat lower electron density ($\log n_e = 9.5$ for the simulation and 9.7 for the
observation) and peak temperature ($\log T_e = 6.04$ for the simulation and 6.11 for the
observation). The dispersion in the DEM also does not match the observation exactly
($\log\sigma_T = 5.24$ and 5.48). Given the approximate nature of the simulations we
consider these discrepancies to be small. The difference in the density comes about
because we have inferred the heating rate from a family of single-loop hydrodynamic
simulations but the emission is actually a composite from several threads. It is likely
that iterating on this solution would yield better agreement with the observations, but
this is unlikely to yield addition physical insights.

\begin{figure}[t!]
\centerline{%
  \includegraphics[clip,scale=0.675]{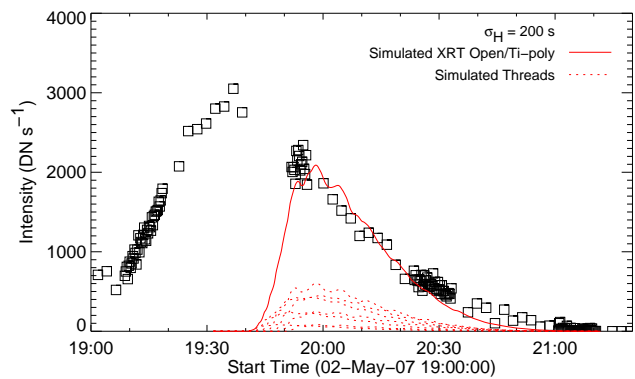}}
\caption{Simulated and observed light curves for the XRT Open/Ti-poly filter combination. }
\label{fig:xrt_sim}
\end{figure}

Finally, we have simulated the expected XRT emission for the Open/Ti-Poly filter
combination using the standard XRT software routine \verb+xrt_t_resp+. The simulated and
observed light curves are shown in Figure~\ref{fig:xrt_sim}. This comparison presents the
greatest challenge to the modeling. The modeled composite intensities, which have been
scaled to match the observations at 20:00 UT, clearly do not extend back in time enough to
cover the entire evolution of the emission in this region. The peak observed emission in
this loop occurs at approximately 19:30 UT, before the simulation has even begun.

Given the diffuse nature of the XRT emission and the difficulty of isolating individual
loops at high temperatures perhaps the simplest explanation may be that the XRT light
curve includes the contributions of many loops in addition to the loop we isolated using
the EIS and \textit{TRACE} data. The identification of individual loops at very high
temperatures with XRT is likely to be hampered by the slow evolution of plasma during the
conductive cooling phase. This is evident by the slow evolution of the threads in the
simulation. In Figures~\ref{fig:eis_sim} and \ref{fig:xrt_sim}, for example, we see that
the threads last for approximately 1 hour at high temperatures. At the lower temperatures,
when radiative losses are much higher, the cooling is dominated by radiation and the
evolution is much faster. In these simulations the threads last only for about 10 minutes
in the \textit{TRACE} 171\,\AA\ bandpass.  This difference suggests that the
differentiated loops seen at lower temperatures, such as those illustrated in
Figure~\ref{fig:eis_color}, would appear as a single structure in XRT. These differences
are also related to the broad temperature response of XRT. The relatively narrow line
emission imaged with EIS and \textit{TRACE} emphasizes small differences in temperature.

Alternatively, the inability of the model to reproduce the observed XRT emission may
reflect inadequacies with the hydrodynamic simulations during the conductive phase of the
cooling. It may be that the heating is not as impulsive as we have assumed. Many previous
studies have suggested a gradual decay in the heating \citep[e.g.,][]{reale2004}. These
differences in the assumed heating may be related to the changes in the topology of the
magnetic field during the evolution of the event.  Observations of post-flare loop arcades
have shown that newly reconnected field lines relax from cusp-shaped to approximately
semi-circular during the early phases of the cooling
\citep{svestka1987,forbes1996,sheeley2004,reeves2008}. The comparisons between the hot and
cool emission shown in Figure~\ref{fig:xrt_trace} clearly suggests that field line
shrinkage is occurring in this event. The cool post-flare loops observed with
\textit{TRACE} are generally observed at the lowest heights of the arcade and do not
overlap with the high temperature XRT emission seen at the top of the arcade. The
implications of field line shrinkage on hydrodynamic simulations has not been
investigated. In general, the conversion of magnetic energy into thermal energy through
the process of magnetic reconnection is not well understood (see, for example,
\citealt{longcope2009}). More detailed analysis of MHD simulations is needed to better
understand the evolution of coronal loops after reconnection \citep[e.g.,][]{linton2006}.

\section{Summary and Discussion}

In this paper we have made use of the unprecedented opportunity to observe evolving
coronal loops in detail. We have used \textit{STEREO}, EIS, \textit{TRACE}, and XRT data
to constrain a multithread model of coronal heating and compare with observations. These
comparisons indicate that it is possible to reproduce the high densities, long lifetimes,
and relatively narrow emission measure distributions inferred from the data so long as the
heating envelope of the heating is sufficiently narrow.

The most challenging comparisons are with XRT, where the model fails to reproduce the
extended lifetime of the emission at high temperatures. It is not clear if this is due to
our inability to isolate narrow loops at high temperatures or to problems with the assumed
envelope on the heating. Recent analysis of the EIS spectral range has identified
\ion{Ca}{14}, \textsc{xv}, \textsc{xvi}, and \textsc{xvii} emission lines that can be used
in the analysis of high temperature plasma
\citep{warren2008c,delzanna2008,watanabe2007}. These lines will provide additional
information on plasma evolution during the conductive phase and have been incorporated
into the latest EIS observing sequences. New active region observations should be
available during the rise of the next solar cycle.

Ultimately our goal is to apply the multithread modeling described here to non-flaring
active region loops. It is encouraging that the simulated EIS differential emission
measure curve derived here is similar to those derived for lower density active region
loops \citep{warren2008b}. It remains to be seen, however, that this models can also
reproduce the loop evolution observed in \textit{TRACE} and XRT. The launch of the
Atmospheric Imaging Assembly (AIA) on the \textit{Solar Dynamics Observatory}
(\textit{SDO}), which will combine full disk imaging, \textit{TRACE}-like spatial
resolution, 10\,s cadences, and multiple filters, will greatly expand the number of useful
active region observations that combine EIS plasma diagnostics and loop evolution.


\acknowledgments The authors would like to thank Jim Klimchuk for helpful discussions on
the time scales for radiative cooling, and Markus Aschwanden for assistance with the
STEREO loop geometry code. Amy Winebarger contributed significant improvements to the
interface to the hydrodynamic code.  Hinode is a Japanese mission developed and launched
by ISAS/JAXA, with NAOJ as domestic partner and NASA and STFC (UK) as international
partners. It is operated by these agencies in co-operation with ESA and NSC (Norway).



\clearpage

\clearpage

\clearpage

\clearpage

\clearpage

\clearpage

\clearpage

\clearpage

\clearpage

\clearpage

\clearpage

\clearpage

\clearpage

\clearpage

\clearpage

\clearpage

\begin{deluxetable}{lrrrr}
\tabletypesize{\footnotesize}
\tablewidth{75mm}
\tablecaption{Observed and Modeled Intensities\tablenotemark{a}}
\tablehead{
\multicolumn{1}{c}{Line} &
\multicolumn{1}{c}{$I_{obs}$} &
\multicolumn{1}{c}{$I_{dem}$} &
\multicolumn{1}{c}{$I_{sim}$} &
\multicolumn{1}{c}{$I_{dem}$}
}
\startdata
\ion{Si}{7} 275.352 & 37.2 & 37.4 & 66.7 & 61.2 \\
\ion{Fe}{10} 184.536 & 190.3 & 194.4 & 339.5 & 342.6 \\
\ion{Fe}{11} 188.216 & 436.9 & 357.9 & 439.7 & 496.7 \\
\ion{Fe}{12} 195.119 & 718.6 & 853.2 & 718.6 & 773.7 \\
\ion{Fe}{13} 202.044 & 595.5 & 545.1 & 301.2 & 277.1 \\
\ion{Fe}{13} 203.826 & 207.9 & 200.3 & 120.7 & 120.9 \\
\ion{Fe}{14} 274.203 & 0.0 & 106.1 & 33.9 & 28.7 \\
\ion{Fe}{15} 284.160 & 0.0 & 296.1 & 36.9 & 31.2 \\
\ion{Fe}{16} 262.984 & 0.0 & 2.71 & 0.1 & 0.1 \\
\enddata
\label{table:ints}
\tablenotetext{a}{Units are erg cm$^{-2}$ s$^{-1}$ sr$^{-1}$. $I_{sim}$ refers
to the simulated intensities presented in Section~\ref{sec:modeling}.}
\end{deluxetable}
 
\clearpage
 
\end{document}